\documentclass[10pt, aps,prc,twocolumn,superscriptaddress,preprintnumbers,
amsmath, 
floatfix,
longbibliography,
nofootinbib
]{revtex4-1}
\usepackage[T1]{fontenc}
\usepackage[utf8x]{inputenc} 
\usepackage{adjustbox}          % Used to adjust figure frames
\usepackage[caption=false]{subfig}
\usepackage{url}
\usepackage{color}
\usepackage{float}
\usepackage[pdftex,colorlinks=true, linkcolor = blue, citecolor=blue,urlcolor=blue, bookmarksnumbered=true, bookmarksopen=true]{hyperref}
\usepackage{longtable}
\usepackage{amsfonts}
\usepackage{dsfont}
\usepackage{wrapfig,bm} 
\usepackage[normalem]{ulem}
\usepackage{MnSymbol}

\newcommand{\beq}{\begin{equation}}
\newcommand{\eeq}{\end{equation}}
\newcommand{\bea}{\begin{eqnarray}}
\newcommand{\eea}{\end{eqnarray}}

\newcommand{\myurl}[1]{\href{https://#1}{\textsf{#1}}}

\begin{document}
\title{Quantum turbulence, superfluidity, non-Markovian dynamics, and wave function thermalization}
  
\author{Aurel Bulgac}
\author{Matthew Kafker}
 \affiliation{Department of Physics,%
 University of Washington, Seattle, WA 98195--1560, USA}
 \author{Ibrahim Abdurrahman}
 \affiliation{ Theoretical Division, Los Alamos National Laboratory, Los Alamos, NM 87545, USA}
 \author{Gabriel Wlaz\l{}owski}
 \affiliation{Faculty of Physics, Warsaw University of Technology, Ulica Koszykowa, 75, 00-662, Warsaw, Poland}
 \affiliation{Department of Physics,%
 University of Washington, Seattle, Washington 98195--1560, USA}
\date{\today}

\begin{abstract}

While quantum turbulence has been addressed both experimentally (predominantly
for superfluid $^4$He and $^3$He) and theoretically, the dynamics of various ensembles of quantized vortices
was followed in time only until the vortices decay into phonons. 
How this ``thermalization'' is achieved is still an unaddressed and thus an unelucidated question.  
The  Unitary Fermi Gas (UFG) is a unique quantum system, which has no classical counterpart and of 
relevance to neutron stars, cold atoms, condensed matter and nuclear many-body systems.   
The non-Markovian evolution of an isolated  UFG is put in evidence and its entire non-equilibrium evolution
can be studied theoretically  within a unified theoretical framework. The initial lattice of quantum vortices 
and anti-vortices evolves through a couple of vortex tangles and excitation of Kelvin waves, 
where vortices cross and reconnect,
until very slowly thermalization sets in.

\end{abstract} 

\preprint{NT@UW-24-08}
\preprint{LA-UR-24-25029}

\maketitle

%%%%%%%%%%%%%%%%%%%%%%%%%%%%%%%%%%%%%%%%%%%%%%%%%%%%%%%%%%%%%%

 \vspace{0.5cm}
 
 In 1955 \textcite{Feynman:1955} conjectured that turbulence can occur in a superfluid,  where viscosity is vanishing at 
 zero temperature, as quantum vortex lines  cross and reconnect virtually in a random manner, 
 and thus the field of quantum turbulence (QT)~\cite{Vinen:2002,Barenghi:2014} came into being. The field of QT has been influenced 
 significantly by the older field of classical turbulence (CT), where in particular Kolgomorov's ideas of energy cascades~\cite{Kolmogorov:1941}
 from the large scale spatial eddies to smaller ones found an extremely  fertile ground. 
 In any fluid, either classical or quantum, there is a natural limiting scale for the energy cascade, dictated by either 
 thermal effects or the atomistic nature of the fluid, see Ref.~\cite{Bandak:2024} and references therein. 
 This last stage, when QT ``dies''  and its ``after life,'' where vortices are not present anymore, has not been addressed in 
 the case of QT. This is a domain of theoretical physics where non-equilibrium dynamics 
 and thermalization of a many-body system~\cite{Neumann:1929,Neumann:2010,Srednicki:1993,Srednicki:1994,
 Horoi:1995,Goldstein:2010,Rigol:2012,Murthy:2023} and entropy production~\cite{Boltzmann:1872,Nordheim:1928,
 Uehling:1933} is typically treated as a Markovian process~\cite{Abrikosov:1963,LL10:1981,Lindblad:1976,Gorini:1976,Griffin:2009}. 
 While the system we have chosen to discuss is not unique, an ensemble of quantum vortices in an isolated 
 UFG, this is a system which both theoretically 
 and experimentally is very well understood~\cite{Zwerger:2011} and of interest to a very large number of physical systems, 
 ranging from neutron stars to nuclei, condensed matter systems, and cold Fermi gases.  
 The same cannot be said about helium 4 or helium 3, which microscopically is discussed mostly within the framework of the 
 Gross-Pitaevskii equation~\cite{Gross:1961,Pitaevskii:1961} and its kinetic extension~\cite{Griffin:2009}, or within a number of phenomenological models.
 Thermalization can be examined for a system stirred in one manner or another, 
 thus not isolated, the time evolution in such a case depends on the manner of coupling between the stirrer and the 
 Fermi superfluid and the theoretical analysis depends on many more variables, and thus the study of an isolated system is clearer.

QT shares similarities with CT, for which dissipation is a critical ingredient. 
CT is studied typically with the Navier-Stokes equations for the vector 
velocity field ${\bm u}$ of an incompressible fluid, which read
\begin{align}
\frac{\partial {\bm u}}{\partial t}+ ( {\bm u} \cdot{\bm \nabla}){\bm u} = \nu {\bm \nabla}^2 {\bm u} 
-\frac{{\bm \nabla }p}{\rho}, \quad {\bm \nabla }\cdot {\bm u}  = 0,
\end{align}
where $\nu$, $p$, and $\rho$ stand for the shear kinematic viscosity, pressure, and 
the matter density respectively. When the viscosity vanishes this equation describes the motion of an ideal fluid, derived by 
Euler in 1775~\cite{Lamb:1975,LL6:1959}. The CT regime occurs when 
the dimensionless Reynolds number ${Re}= uL/\nu$ is large or $\nu\rightarrow 0$, where $L$ 
is the characteristic length of the flow. The solutions of time-dependent partial differential 
equations change qualitatively when the highest order of the spatial derivatives changes from second to first~\cite{Whitham:1977}. 
The Navier-Stokes equations can be derived from the 
Boltzmann transport equation~\cite{Boltzmann:1872,Nordheim:1928,Uehling:1933,LL10:1981}, 
which consists of the free transport part, which describes the independent mean 
field particle motion, and the collision integral, responsible for local equilibration. 
The Boltzmann equation is strictly valid for dilute systems, when collisions are assumed to happen ``instantaneously 
and at a particular point  in space~\cite{LL10:1981}.'' The evolution of the distribution function is described with the Boltzmann equation 
``over times long compared with the duration of collisions'' and ``over distances 
large compared with the size of the region in which a collision takes place.''~\cite{LL10:1981} 
Under such conditions, the mean free path exceeds the average separation between particles.
These assumptions~\cite{LL10:1981} imply that different collisions are uncorrelated in time and space, 
and hence have a Markovian character, in a manner similar to the Langevin noise in Brownian motion.

Consider the simple example of one or more ``fluids'' made up of dust particles, thus non-interacting, and with sizes negligible in comparison 
to their average separation, and each ``fluid'' with its own velocity probability distribution, in the absence of any external fields.
All elementary constituents follow trajectories with a constant velocity. Each such ``fluid''  satisfies the Euler equations, and
moreover, each dust ``fluid'' satisfies the Navier-Stokes equations independently. 
This aspect clearly demonstrates that the limit $\nu\rightarrow 0$   is undefined, as the fractions of different ``fluids'' can be arbitrary at any point in space. 
One can have a very strange situation, for example, two or more  interpenetrating ``dust rivers,''  which do not affect each other, 
in which dust particles in each ``river'' follow their independent trajectories, which can even cross each other. 
Any small interaction between particles would ``eliminate'' crossings, leading to what in quantum mechanics is often referred to as avoided  level crossings. 
One can define an average density and velocity at each point in space, but the time evolution of such a mixture cannot be described by Euler equations. 
No well-defined continuous limit from a fluid with interactions ($\nu \neq 0$) to a ideal fluid exists, which may elucidate why 
turbulent motion appears so formally complex, see also Ref.~\cite{Bandak:2024} and earlier references therein for related situations.  

In the case of superfluids (such as liquid helium 3 and 4, 
neutrons and protons in neutron stars, nuclear systems, cold atom systems) 
the interactions between the elementary constituents are strong,  
unlike in an ideal fluid, which is assumed 
to be in local equilibrium at all times in the case of ideal gases.  
However, at $T=0$  the viscosity of 
the superfluids vanishes, and according to the Feynman conjecture~\cite{Feynman:1955,Vinen:2002,Barenghi:2014,QT:2014} quantized vortices can cross and 
reconnect and generate a ``turbulent''  evolution of  the superfluid. 
One can infer that the main difference between classical and QT, 
which according to \textcite{Feynman:1955} happens when quantized vortices cross and reconnect,  
is in the strength of the interactions between 
the elementary components of the fluid. 
Since at finite temperatures, but below the critical temperature, 
a superfluid has both normal and superfluid 
components~\cite{Tisza:1938,Landau:1941} the turbulent motion can have mixed 
characteristics  of both quantum and classical fluids, due to the ``friction'' between 
the normal and superfluid components and the excitation of Kelvin waves~\cite{Vinen:2002,Barenghi:2014,QT:2014}. 

The UFG~\cite{Braaten:2011,Bulgac:2011a,Castin:2012,Zwerger:2011} is a remarkable strongly
interacting quantum Fermi system, with a very large pairing gap and a large critical temperature~\cite{Bulgac:2006a,Magierski:2011}. 
The fermion-fermion interaction is characterized by an infinite scattering length and a zero-range interaction~\cite{Bethe:1949}, 
particles interact only in the $s$-wave, the scattering phase shift is $\delta_0(k) \equiv \pi/2$, and 
the two-fermion scattering cross section reaches its maximum allowed by unitarity $\sigma(k)=4\pi/k^2$, where $k$ is their relative wave vector. 
The only dimensional parameter that determines the 
properties of the UFG is the average inter-particle separation~\cite{Zwerger:2011}. (We use units $\hbar=m=1$.)
The UFG is a pure quantum system with no classic limit. It has the remarkable behavior at any energy, 
both in and out of equilibrium,  that the single-particle occupation probabilities 
$n({\bm k}) \rightarrow C/k^4$, in the limit $k\rightarrow \infty$~\cite{Tan:2008a,Tan:2008b,Tan:2008c}. 
Thus, the total particle number $\int d^3k \,n({\bm k})$ converges very slowly, a patently non-Maxwellian behavior.   

The real-time dynamics of the UFG is described within an extension of the Time-Dependent Density Functional Theory (TDDFT) 
to fermionic superfluids, the 
(Time-Dependent) Superfluid Local Density Approximation 
((TD)SLDA)~\cite{Bulgac:2007,Bulgac:2013a,Bulgac:2016x,Bulgac:2019}, which by design looks like the 
Time-Dependent-Hartree-Fock-Bogoliubov approximation. 
It is trivial to show that the TDSLDA framework is fully equivalent to the description of a non-equilibrium evolution of a 
fermionic superfluid with 
Gorkov equations ~\cite{Gorkov:1958,Abrikosov:1963,Gorkov:2010}, 
or, alternatively, with a generalized fully quantum Boltzmann kinetic equation~\cite{Bulgac:2022}. 
While in the (semi-)classical Boltzmann equation~\cite{Boltzmann:1872,Nordheim:1928,Uehling:1933,LL10:1981} one can introduce a phase-space distribution,
in the deep quantum regime only the canonical/natural occupation probabilities can be meaningfully defined for a 
time-dependent system~\cite{Coleman:1963,Lowdin:1956,Lowdin:1956a,Davidson:1972,Bulgac:2023}. 
Unlike in the (semi-)classical Boltzmann equation, in TDSLDA the kinetic evolution of a many-body system can 
describe interference effects and entanglement~\cite{Bulgac:2022c,Bulgac:2023,Bulgac:2023c}.
For the justification, validity, verification, and structure of TDSLDA  the reader can
consult Refs.~\cite{Bulgac:2007,Bulgac:2013a,Bulgac:2016x,Bulgac:2019,Shi:2020} and studies referred therein. 
The energy density functional in TDSLDA unfortunately cannot be directly linked with the particle interactions in the 
Schr{\" o}dinger equation, except by ``fitting'' its dimensionless parameters and reproducing accurate Quantum Monte Carlo 
data as described in Refs.~\cite{Bulgac:2007,Bulgac:2011a}. In this manner the infinite scattering length $a$ 
and the zero-rage interaction radius $r_0$ become a set of finite dimensional parameters
$\alpha, \beta,$ and $\gamma$~\cite{Bulgac:2007,Bulgac:2011a}, similar to the running coupling 
constants in  quantum field theory. The UFG is indeed a strongly interacting system with an estimated mean free path 
of a quasiparticle around the Fermi level given by $\lambda \approx  1/(\sigma (k)n_0) \approx  {\cal O}(n_0^{-1/3})$, where $n_0=k_F^3/3\pi^2$ is 
the average number density. This is true only if the Pauli effect is not 
accounted for, which is crucial, see Refs.~\cite{Nordheim:1928,Uehling:1933}. The vacuum fermion-fermion scattering cross section 
$\sigma(k)=4\pi/k^2$ is strongly renormalized in the medium and it can lead to strong induced $p$-wave 
interactions and even a $p$-wave pairing gap between fermions with 
identical spins~\cite{Bulgac:2006,Bulgac:2009}, similar in spirit to the proton-neutron pairing in nuclear system~\cite{Bulgac:2022}. 
The $p$-wave correlations effects have not yet been incorporated into TDSLDA studies.

The numerical simulations described below were performed on supercomputers Frontier and Summit at Oak Ridge 
National Laboratory, using code W-SLDA Toolkit~\cite{WSLDAToolkit}, specifically designed 
to describe the time evolution of a UFG accurately. The capabilities of the framework in 
describing complex phenomena involving quantum vortices was demonstrated in 
works~\cite{Bulgac:2011,Bulgac:2011s,Bulgac:2011a,Bulgac:2014,Wlazlowski:2015,Wlazlowski:2018,Hossain:2022,Barresi:2023,
Bulgac:2022c,Wlazlowski:2024,Bulgac:2011,Bulgac:2011s}. The code solves the TDSLDA equations on 
spatial 3D lattice of size $N_x \times N_y \times N_z$. Similar codes, but for nuclear systems, were described in~\cite{Shi:2020,pecak2024}.  
The UFG dynamics was simulated in a cubic box ($N_x = N_y = N_z=32$) of size $L^3 = (N_x l)^3$ with 
periodic boundary conditions and a finite lattice constant $l$. 
In this respect, our numerical implementation is similar to Lattice Quantum Chromodynamics and 
the only numerical approximation is in using a finite time-step, with well-controlled errors ${\cal O}(\Delta t^5)$.    
With this code, we evolved the $2N_x^3$ = 65,536 complex, nonlinear, coupled, 3D+time, PDEs 
for up to $3.75\times 10^6$ time-steps with relative numerical errors of $10^{-6}$ and $10^{-8}$ respectively for 
the conservation of the total energy and of the total particle number. 

The dynamics of the quantized linear and ring vortices  in the UFG, their crossing, 
and their reconnections were studied previously in Ref.~\cite{Bulgac:2011,Bulgac:2011s,Bulgac:2014,Wlazlowski:2015}.
The vortex dynamics is qualitatively similar to Feynman's conjecture~\cite{Feynman:1955}.  
In the present study we start with an ensemble of 12 linear vortices and anti-vortices, 
as an initial constrained state of an unpolarized UFG, at $T=0$~\cite{Bulgac:2022c}, with vanishing total circulation. 
As the ensemble of vortices evolves in time vortices start bending, crossing 
and reconnecting, while emitting phonons,  see movie in the Supplemental Material (SM)~\cite{supplemental_material}). 
Their total linear length shortens with time, until eventually no remnants of the 
initial vortex configuration survives, and the entire system reaches a ``thermal'' state. 
We have simulated  an unpolarized UFG  with the total number of particles 
$N=N_\uparrow+N_\downarrow\approx 1000$, 
where $N_\uparrow$ and $N_\downarrow$ are the number of fermions with spin up and down. 
The average spatial number density is approximately $n_0=N/V\approx 0.03$ for  
the lattice constant  $l=1$.  The size of the many-body 
Fock space for an unpolarized  UFG ($N_\uparrow=N_\downarrow$) in such a cubic box 
for $N$ fermions is~\cite{Bulgac:2023} $2^{(N_xN_yN_z)^2}$ where $(N_xN_yN_z)^2\approx 1.07\times10^9$.
Similar results, 
but for shorter trajectories, were obtained for lattice sizes $l=0.8, 0.5$ and correspondingly higher momenta cutoffs $k_{cut}=\pi/l$. 
One can use Bethe's 
approximate estimate of the energy level density of a fermion system~\cite{Bethe:1936}
\begin{align}
\rho(E)\propto \exp[2\sqrt{aE}]\propto \exp(S(E))\approx {\cal O}[\exp(cN)],\label{eq:3}
\end{align}
where $S$ is the thermodynamic entropy, $c \approx{\cal O}(1)$ and $N\approx 1,000$. $\rho(E)\approx {\cal O}(10^{100})$ in our case 
gives an idea of how many states the system 
has to visit in an energy shell $\Delta E$ in order to satisfy Gibbs's equality between 
the infinite time average and the phase space average of any observable. 

 %%%%%%%%%%%%%%%%%%%%%%%%%%%%%%%%%%%%%%%%%%%%%%%%%%
\begin{figure}
\includegraphics[width=1.0\columnwidth]{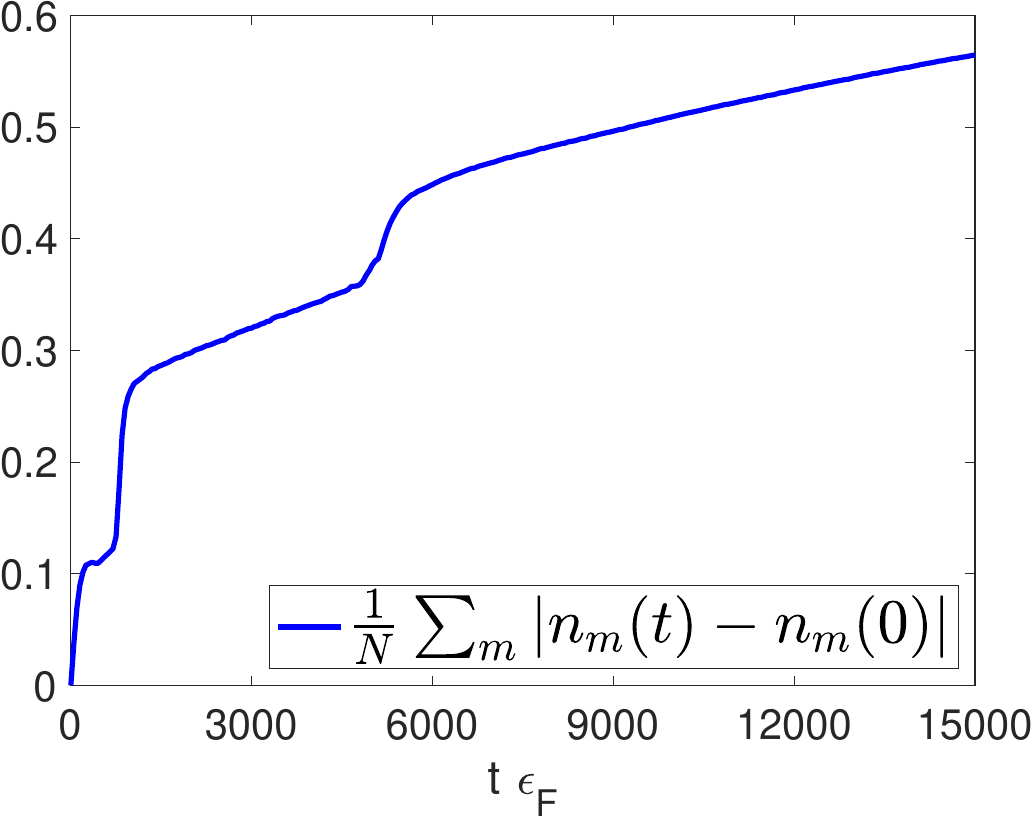}  
\caption{ \label{fig:DensityC}  
The fractional sum $0\le \sigma_1(t)\le 2$ of the absolute differences between
single-particle occupation probabilities $n_m(t)$ at the initial time and at the final time $t$. See the video ufg32 in supplemental material~\cite{supplemental_material}. }
\end{figure}  

%  
 %%%%%%%%%%%%%%%%%%%%%%%%%%%%%%%%%%%%%%%%%%%%%%%%%% 
 
\begin{figure}
\includegraphics[width=1.0\columnwidth]{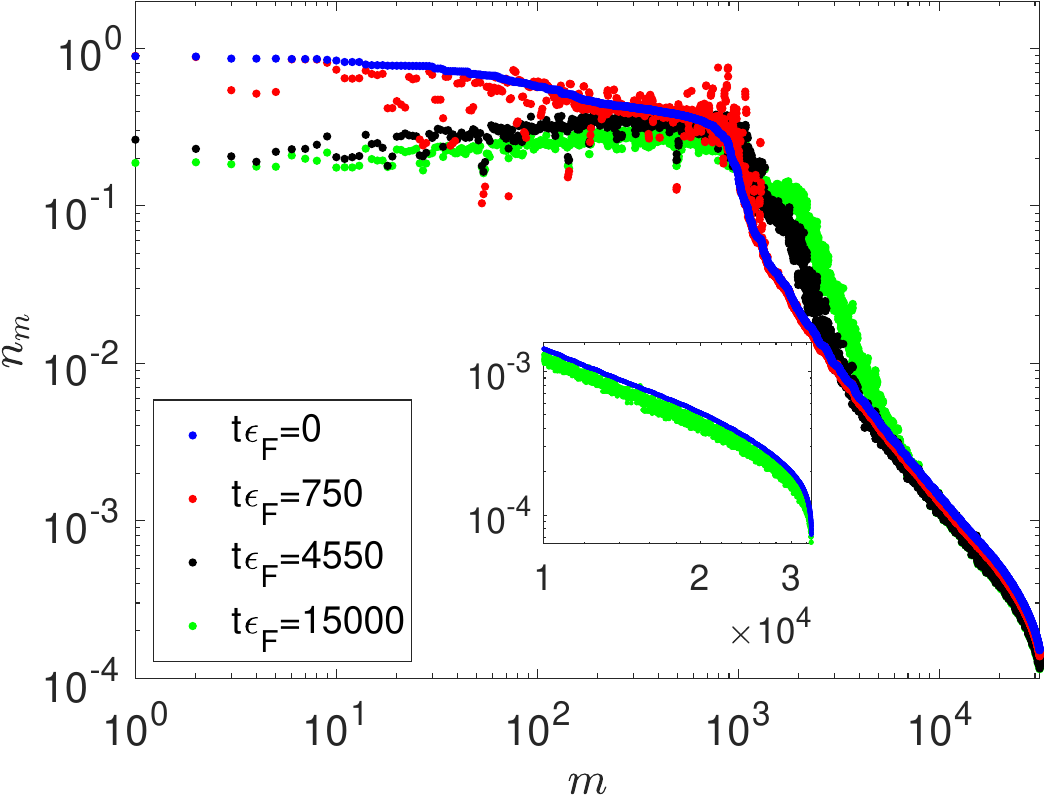}  
\caption{ \label{fig:SpectrumEvolution} 
The  time-dependent occupation probabilities $n_m(t)=v^2_m(t)$ at several times during the evolution. 
Due to the $s$-wave collisions the single-particle occupation probabilities in the 
final ``thermalized'' state are drastically redistributed. The states are ordered in decreasing order of the initial occupation probability $n_m(0)$. 
Since $\sum_m n_m(t)=N_\uparrow=N_\downarrow$ the initial high occupation probability quasiparticle states are 
depopulated at the expense of quasiparticle states with higher energies, but smaller 
than the upper cutoff energy $\epsilon_{cut} \approx {\cal O}( \pi^2/2l^2)$, 
see inset. 
}
\end{figure}   
 
%  
 %%%%%%%%%%%%%%%%%%%%%%%%%%%%%%%%%%%%%%%%%%%%%%%%%%

\begin{figure}[h]
\includegraphics[width=1.0\columnwidth]{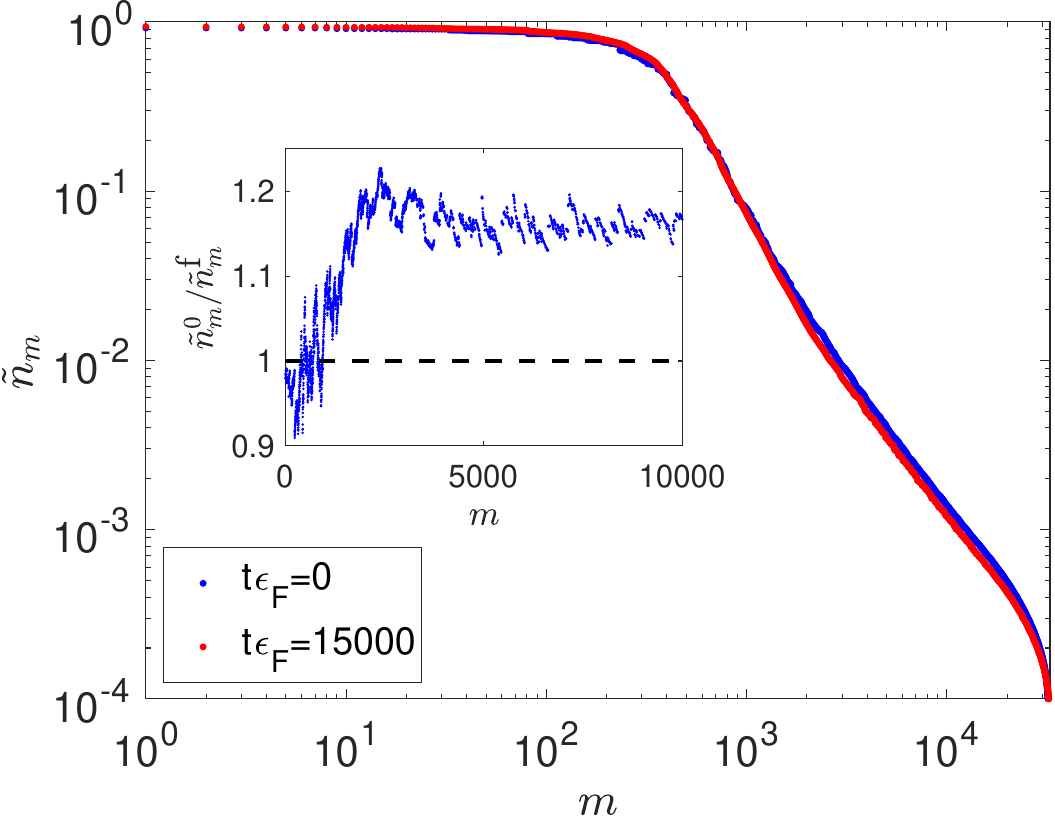}  
\caption{ \label{fig:CanOccs} 
The initial and final canonical occupation probabilities $\tilde{n}_m(t) = \tilde{v}_m^2(t)$, ordered by magnitude, which are the eigenvalues of the correspond 
instantaneous one-body density matrix, see Ref.~\cite{Bulgac:2022,Bulgac:2024}.  In the inset we show the ratio of the initial 
over final canonical occupation probabilities $\tilde{n}_m^{0}/\tilde{n}_m^f$. Since canonical occupation probabilities change with 
time there is no one-to-one correlation between their subscripts $m$.  }
\end{figure}

\begin{figure}
\includegraphics[width=1.0\columnwidth]{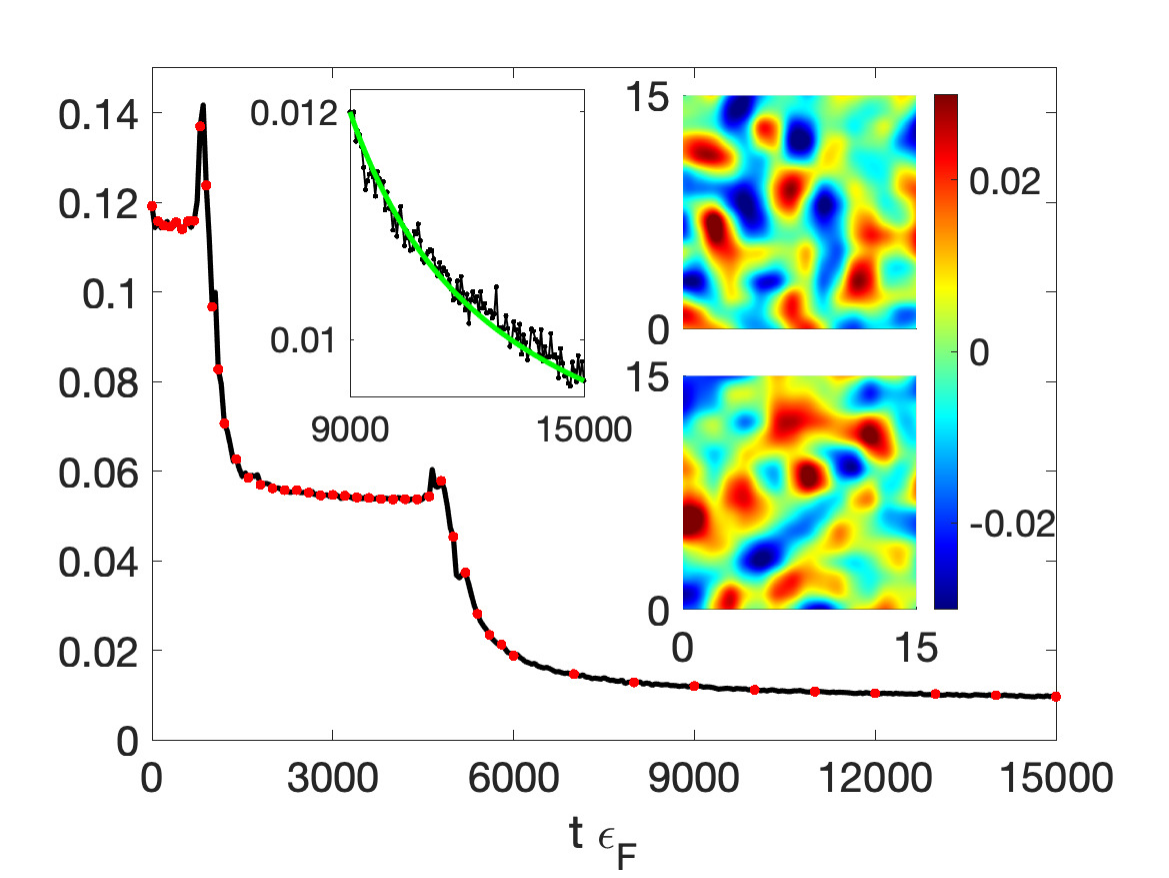}  
\caption{ \label{fig:DensityF} The average fluctuation of the integrated number density $\int_V d^3r | n({\bm r},t)/n_0-1|/V$ (black line) as a 
function of time for the raw number density in TDDFT with pairing correlations and for the particle-number projected density at each 
time plotted at selected times (red points)~\cite{Bulgac:2024}, shows almost perfect agreement as expected in 
TDDFT with~\cite{Oliveira:1988} or without pairing correlations~\cite{Runge:1984} present. 
Here $n_0 = N/V$ is the average particle number density and $V$ is the volume of the cubic simulation box. 
In the upper left inset we show $\int_V d^3r | n({\bm r},t)/n_0-1|/V$  in the quasi-homogeneous 
stage for $t\epsilon_F>9000$ compared to a numerical fit (green line), see Eq.~\eqref{eq:fit}. 
In the right two insets we show the instantaneous number density fluctuations $n(x,0,z,t)/n_0-1$ at times $t\epsilon_F = 10000$ 
and $10050$.   See the second movie in SM~\cite{supplemental_material}.}
\end{figure} 

\begin{figure}
\includegraphics[width=1.0\columnwidth]{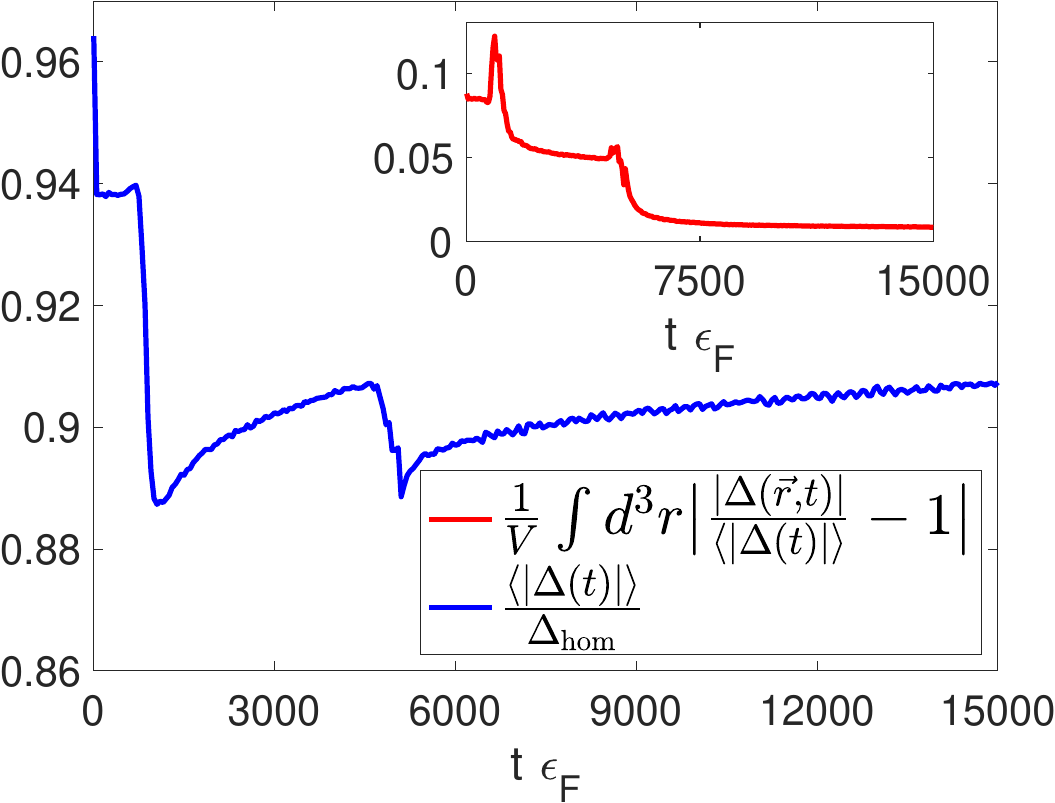}  
\caption{ \label{fig:PairingGap}
The time evolution of the average of absolute value of the pairing gap, divided 
by its value in the homogeneous UFG at the same number density. The inset shows the average fluctuation of this quantity. 
}
\end{figure}  

In the deep quantum regime, the only meaningful counterpart to the (semi-)classical phase-space distribution typically 
used in kinetic equations, are the single-particle occupation probabilities $n_m(t)$. 
Fig.~\ref{fig:DensityC} illustrates the dramatic change with time of the single-particle occupation probabilities, 
using the fractional occupation probability sum
\begin{align}
\sigma_1(t)& =\frac{\sum_m|n_m(t)-n_m(0)|}{N}, \quad 0\leq \sigma_1(t)\leq 2.
\end{align}
For a gas, which almost always expands to occupy the entire volume  in most cases and one expects that $\sigma_1(t)\leq 1$, 
unlike in the case of a liquid or solid. 
In the case of a time-dependent Hartree-Fock simulation
the occupation probabilities will remain either 1 or 0 for all times. In the presence of $s-$wave interactions, 
the initial single-particle occupation probabilities  are those of a fermion  system with pairing correlations 
in the presence of quantized vortices, see blue dots in Fig.~\ref{fig:SpectrumEvolution}. For states with relatively small quasi-particle energies 
the occupation probabilities are somewhat similar to the Bardeen-Cooper-Schrieffer 
(BCS)~\cite{Bardeen:1957} single-particle distribution. For zero-range interactions and arbitrary scattering length 
the occupation probabilities have the universal large momentum tail $k$-behavior of the momentum distribution
$n(k,t) \approx C/k^4$ for any many-fermion state~\cite{Tan:2008a,Tan:2008b,Tan:2008c,Zwerger:2011,Castin:2012,Braaten:2011}, 
see Fig.~\ref{fig:CanOccs} for the canonical occupation probabilities. 
For $k\gg k_F$ the energy of single-particle
states with high $m$-values is dominated by the kinetic energy alone and a power law behavior 
is apparent in these log-log plots for $m > 3,000$ at all times, see Refs.~\cite{Bulgac:2022,Bulgac:2023,Bulgac:2024}. 
For large momenta $k$ the index $m \approx 4\pi k^3/3$ (as in the case of non-interacting fermions) and $\tilde{n}_m(t) \approx \tilde{C}/m^{4/3}$.

In the case of a time-dependent system one has to introduce the canonical occupation 
probabilities at each time step, see Ref.~\cite{Bulgac:2023}, since 
particle occupation probabilities $n_m(t)$ are not uniquely defined.  The canonical occupation probabilities are 
gauge invariant and instantatenous eigenstates of the 
time-dependent one-body density matrix~\cite{Bulgac:2024}. 
Then the time-dependent many-body state acquires the BCS form
\begin{align}
&|\Phi(t)\rangle = \prod_m [\tilde{u}_m(t)+\tilde{v}_m(t) a^\dagger_m(t) a^\dagger_{\overline{m}}(t)]|0\rangle,\\
&\alpha_{m,\overline{m}}(t) = \tilde{u}_m(t)a_{m,\overline{m}}(t)\mp \tilde{v}_m(t)a^\dagger_{{\overline{m},m}},
\end{align}
where $\alpha_{ m, \overline{m}}(t) |\Phi(t)\rangle=0$, $\tilde{u}
^2_m(t)+\tilde{v}^2_m(t)=1,$
$0\le \tilde{u}_m(t),$ $\tilde{v}_m(t)\le 1$, $N=2\sum_m \tilde{v}^2_m(t)$, and  $a^\dagger_m(t),$ $a^\dagger_{\overline{m}}(t)$ 
are time-dependent  canonical creation fermionic operators~\cite{Bulgac:2023}.
Surprisingly, the initial and final canonical occupation probabilities
$\tilde{n}_m(t)$ are very similar, see Fig.~\ref{fig:CanOccs}. 

In (semi-)classical Boltzmann kinetic approaches~\cite{Boltzmann:1872,Nordheim:1928,Uehling:1933,LL10:1981} a collision integral 
is added to the equations for the phase-space distribution, which is derived at each time step by using the hypothesis
that the system  attained a ``molecular'' equilibrium/chaos, and then evaluating the change 
in the phase-space distribution at each time step using perturbation theory. 
This implies 
that the dynamics is slow enough that after each particle collision the system equilibrates. In this respect that collision integral plays the same role 
as the noise in the Langevin equation, which describes the Brownian motion. 
Consequently, one expects in a pure kinetic approach that the redistribution of the occupation 
probabilities should have a similar Markovian character, and that the departure of the phase-space distribution from its initial values as a 
function of time should be proportional to $\sqrt{t}$. 
However, the time evolution of the quantity $\sigma_1(t)$~\cite{Bulgac:2022,Bulgac:2023,Bulgac:2024} illustrated in 
Fig.~\ref{fig:DensityC}  shows a clear piecewise linear dependence on time, 
depending on the character of the system at different stages during the evolution.
We prefer to illustrate the non-Markovian behavior by focusing on 
$0\le \sigma_1(t)\le 2$  as $\sigma_1(t)$ is a 
transparent measure of the fractional change in all occupied single-particle states.  
The quantity, $\sigma_1(t)$, reaches its lower limit in the case when the occupation probabilities do not change in time, 
e.g. in a time-dependent Hartree-Fock problem, when particle collisions are absent. 
In a UFG there is only one scale for the single-particle energies, as the mean field and the pairing field/collision term are of the order 
of the Fermi energy of  free Fermi gas $\epsilon_F=k_F^2/2$~\cite{Zwerger:2011}, and those are the only quantities 
expected to control the occupation probabilities re-distributions. 
Moreover, unlike in the (semi-)classical Boltzmann evolution, the ``collisions'' are not isolated for an UFG either in time or in space, 
the system is not dilute in this deep quantum limit, since scattering cross section between two fermions in UFG $\sigma =4\pi/k^2$ 
is comparable to the square of the average separation between particles $n({\bf r},t)^{-2/3}\approx \pi^2/k_F^2$, 
and the quantum ``jumps''  are thus correlated and particles are entangled. The equivalent of 
$f({\bm r},{\bm p},t)\rightleftarrows 1-f({\bm r}',{\bm p}',t)$ transition in the semi-classical Boltzmann equation~\cite{Nordheim:1928,Uehling:1933} 
is the action of the pairing gap 
$u_m({\bm r},\sigma,t) \rightarrow \Delta({\bm r},t)v_m({\bm r},-\sigma,t)$ or   
$v_m({\bm r},-\sigma,t) \rightarrow -\Delta^*({\bm r},t)u_m({\bm r},\sigma,t)$ in the TDSLDA equations~\cite{Bulgac:2022},
identical to the collisions described by the Gorkov equations~\cite{Gorkov:1958,Gorkov:2010,Abrikosov:1963}. 
In a UFG $u_m({\bm r},\sigma,t)$, $v_m({\bm r},\sigma,t)$, and $\Delta({\bm r},t)$ are fully delocalized in space.

For very small times $t\epsilon_F <700$ the linear vortices start deforming slightly, but still do not cross~\cite{supplemental_material}. 
In the next stage, for times $700< t\epsilon_F < 1200$, 
the vortices form a tangle, cross, reconnect, and their overall length shortens. The dynamics result in the 
formation of a state consisting of four aligned vortices. They persist up to time  $t\epsilon_F\approx 4300$, 
after which they start to reconnect again and finally annihilate. For times $t\epsilon_F > 5000$ 
only local fluctuations of the number densities survive in the system, see SM~\cite{supplemental_material}.
It is not surprising that the rate of change of $\sigma_1(t)$ varies, depending on the presence, density, and character of vortices in the system.  

After crossing and reconnections of the vortex lines, and ``thermalization'' of the system for times $t\epsilon_F > 6,000$, 
see Fig.~\ref{fig:DensityF}, the system becomes quasi-homogeneous, apart from small 
quantum fluctuations of the  number density  $\approx 0.01 n_0$. It is worth stressing, 
that even though in the presence of pairing correlations the particle number is conserved only on average, the raw time-dependent 
TDDFT number density and the particle-projected time-dependent number density~\cite{Bulgac:2024} show essentially a 
perfect agreement with each other. Judging by the final occupation probabilities in 
Fig.~\ref{fig:SpectrumEvolution} (green dots) one would naively expect the final state would be characterized by a very high temperature. 
In the ``thermal'' regime ($t\epsilon_F > 6,000$) 
the average spatial fluctuations of $n({\bm r},t)$ are of the order of $1\%$, see Fig.~\ref{fig:DensityF}, and a similar order for the pairing gap, see 
Fig.~\ref{fig:PairingGap}. 
Furthermore, the pairing gap in the initial state, where quantized vortices are present,  and in the final state, where 
the quantized vortices have ``melted'' into phonons, are  surprisingly similar as well, the spatially averaged absolute value of the pairing gap 
changed by less than $\approx 5\%$, see Fig.~\ref{fig:PairingGap}, and in the final state these fluctuations
are at the level of $\approx 1\%$. The entire system is now at a finite ``effective'' temperature $T$, 
below the critical temperature $T_c\approx0.16 \epsilon_F$~\cite{Ku:2012}.  Furthermore, UFG is 
characterized by relatively small amplitude thermal quantum fluctuations of the number density and of the pairing gap, 
see Figs.~\ref{fig:DensityF} and \ref{fig:PairingGap}. While in the initial state there are currents, 
due to the presence of quantized vortices,  which contribute about 10\% to the total energy of the system, 
in the ``thermalized'' state the currents contribute only about 0.01\% to the total energy and the system is basically at stand still.

It is instructive to compare the time evolution of the average fluctuation of the number density illustrated 
in  Fig.~\ref{fig:DensityF} to the Eigenstate Thermalization Hypothesis (ETH) prediction~\cite{Srednicki:1994},
which  concludes ``that a generic initial state will approach thermal equilibrium at least as fast as
$t_{therm} \approx {\cal O}(\hbar/\Delta E)$, where $\Delta E$ is the uncertainty in the total energy of the gas.''
We construct the simple fit to $\delta [n(\vec{r},t)]$  
\begin{align}
 &\delta [n(\vec{r},t)]= \frac{1}{ V} \int d^3r \Bigg |  \frac{ n(\vec{r},t) }{n_0}-1 \Bigg |  
\approx \delta n_\infty +\frac{t_{UFG}}{t-t_0}. \label{eq:fit}
\end{align}
$\delta n_\infty$ is the average value of the time-dependent quantum number-density fluctuations 
in the limit $t\rightarrow \infty$, when the system is fully thermalized. 
$\delta [n(\vec{r},t)]$ is an observable and in principle can be measured in cold atomic clouds experiments~\cite{Bakr:2009}.
%%%%%%%%%%%%%%%%%%%%%%%%%%%%%%%%%%%%%%%%%%%%%%%%%%%%%%%%%%%%%%%%
\begin{figure}[h]
\includegraphics[width=1.0\columnwidth]{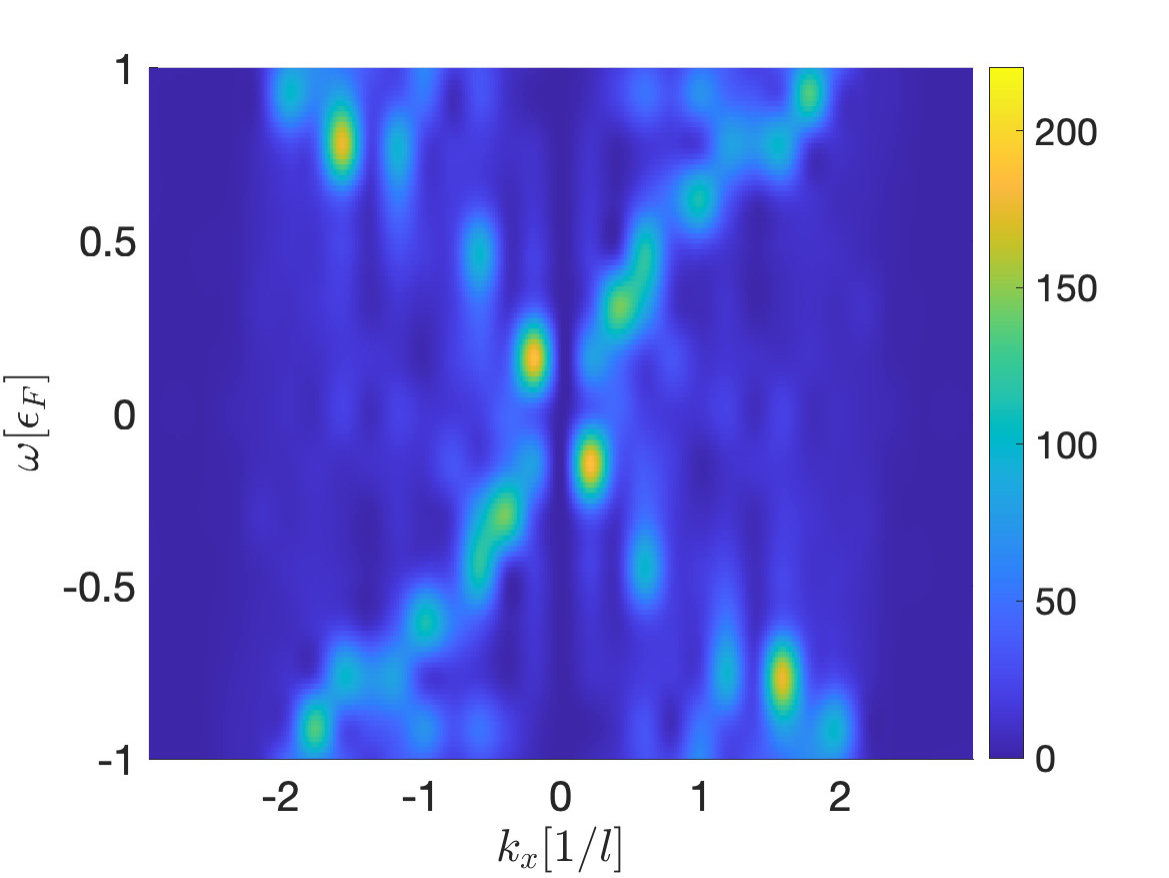} 
\caption{\label{fig:SpectralFunction} The profile of the space-time Fourier transform of density fluctuations 
$ S(\omega,k_{x,y,z})= \int_V d^3r \int _T^{T+50} dt [n({\bm r},t)/n_0-1] \exp(i{\bm k}\cdot  {\bm r}-i\omega t)$, 
where $V$ is the simulation box volume, 
$T = 15,000\epsilon_F$, and ${\bm k}$ was chosen parallel to the axis $Ox$. Other choices of ${\bm k}$ lead to very similar results.  
}
\end{figure} 
%%%%%%%%%%%%%%%%%%%%%%%%%%%%%%%%%%%%%%%%%%%%%%%%%%%%%%%%%%%%%%%%
 In lower upper left inset of Fig. \ref{fig:DensityF} we have chosen for the fit $t_0\,\epsilon_F =5050$ , 
 when the UFG is in the quasi-homogeneous stage of the time evolution, and obtained
 for the relaxation times in the case of UFG and ETH respectively we obtain the values
 \begin{align}
 &t_{ETH}\, \epsilon_F \approx  \frac{1}{\Delta E} ={\cal O}(0.1)    \ll t_{UFG}\,\epsilon_F={\cal O}(10),
 \end{align}
 where $\Delta E \approx \mu\Delta N(t)\approx  15\ldots 10\, \epsilon_F$ for $t_0\epsilon_F  = 5055,\ldots, 6050$,  $\mu$ is the UFG chemical potential and 
 $\Delta N(t) = 2 \sqrt{\sum_m\tilde{n}_m(t)(1-\tilde{n}_m(t))}\approx 30$ is the particle number variance in the TDSLDA  simulation.   
 In TDSLDA simulations the total energy of an isolated system is conserved and $\Delta E \equiv 0$, however, the particle number 
 is conserved only on average, as in TDSLDA one evolves an ensemble of many-body states with different particle numbers. 
 As Fig.~\ref{fig:DensityF} and Ref.~\cite{Bulgac:2024} show 
 the difference between the raw time-dependent TDDFT $n({\bm r},t)$ and the number-projected one are small enough and could be ignored.
 $t_{UFG}$ does not vary dramatically for other choices $t_0\epsilon_F> 5000$.
 In  the case of UFG it appears that the thermalization time is longer by orders of magnitude 
 than predicted rate for ETH~\cite{Srednicki:1994}. It is expected that for 
 $t\epsilon_F > 9000$ only phonons with dispersion $\omega^2 = k^2c_0^2$ survives, 
 where  $c_0 =k_F \sqrt{\xi/3}$
 is the sound velocity and $\xi$ is the Bertsch parameter~\cite{Zwerger:2011},  see Fig. \ref{fig:SpectralFunction}. 
 As the system did not reach yet a full thermalization, the expected $\omega = \pm kc_0$ is observed only on average 
 for the limited time interval $t\,\epsilon_F \in (15000, 15050)$.  
The ``wave function thermalization'' process discussed in this work is much slower 
than that discussed previously in literature~\cite{Srednicki:1994}.  The UFG is a unique type of quantum system with no classical limit, 
and Berry's conjecture does not apply~\cite{Berry:1977,Berry:1991,Srednicki:1999}. 
\\
%%%%%%%%%%%%%%%%%%%%%%%%%%%%%%%%%%%%%%%%%%%%%%%%%%

A discussion with E. K. U. Gross of this work is greatly appreciated.
AB designed the study, analyzed and interpreted the results 
of the simulations, and wrote the first draft of the manuscript. GW 
wrote the code used in these simulations and MK, IA, and GW 
performed the numerical simulations. All authors participated in
writing of the final version of the manuscript.
The funding for AB from the Office of Science, Grant No. 
DE-FG02-97ER41014 and also the partial support provided by NNSA 
cooperative Agreement DE-NA0003841 is greatly appreciated. 
MK was supported by NNSA cooperative Agreement DE-NA0003841.
 The work of I.A. was supported by the U.S.
Department of Energy through the Los Alamos National
Laboratory. The Los Alamos National Laboratory is operated
by Triad National Security, LLC, for the National Nuclear
Security Administration of the U.S. Department of
Energy Contract No. 89233218CNA000001. 
 GW  was supported by the Polish National Science Center (NCN) 
under Contract No. UMO-2022/45/B/ST2/00358.
 This research used resources of the Oak
Ridge Leadership Computing Facility, which is a U.S. DOE Office of
Science User Facility supported under Contract No. DE-AC05-00OR22725. 

%
 % These are needed to avoid a babel error.
\providecommand{\selectlanguage}[1]{}
\renewcommand{\selectlanguage}[1]{}

\bibliography{local_fission1}

\end{document}